**Title:** Vitamin K content of cheese, yoghurt and meat products in Australia

**Authors:** Eleanor Dunlop[a], Jette Jakobsen[b], Marie Bagge Jensen[b], Jayashree Arcot[c], Liang Qiao[d], Judy Cunningham[a], and Lucinda J Black[a,e*]

**Author affiliations:**

[a] Curtin School of Population Health, Curtin University, Kent Street, Bentley WA 6102, Australia. eleanor.dunlop@curtin.edu.au; judyc121@gmail.com; lucinda.black@curtin.edu.au

[b] Research Group for Bioactives – Analysis and Application, National Food Institute, Technical University of Denmark, Lyngby, Denmark; jeja@food.dtu.dk; mabaj@food.dtu.dk

[c] School of Chemical Engineering, University of New South Wales, Sydney, Australia; j.arcot@unsw.edu.au

[d] Storr Liver Centre, The Westmead Institute for Medical Research, Westmead, Australia; liang.qiao@sydney.edu.au

[e] Curtin Health Innovation Research Institute (CHIRI), Curtin University, Perth, Australia; lucinda.black@curtin.edu.au

**\*Corresponding author:** Lucinda J Black, Curtin School of Population Health, Curtin University, GPO Box U1987, Perth, WA 6845, Australia. lucinda.black@curtin.edu.au. Tel.: +61 8 9266 2523. Fax: +61 8 9266 2958.

**Abstract (max 150 words)**

Vitamin K is vital for normal blood coagulation, and may influence bone, neurological and vascular health. Data on the vitamin K content of Australian foods are limited, preventing estimation of vitamin K intakes in the Australian population. We measured phylloquinone (PK) and menaquinone (MK) -4 to -10 in cheese, yoghurt and meat products (48 composite samples from 288 primary samples) by liquid chromatography with electrospray ionisation-tandem mass spectrometry. At least one K vitamer was found in every sample. The greatest mean concentrations of PK, MK-4 and MK-9 were found in lamb liver, chicken leg meat and Cheddar cheese, respectively. Cheddar cheese and cream cheese contained MK-5. MK-8 was found in Cheddar cheese only. As the K vitamer profile and concentrations appear to vary considerably by geographical location, Australia needs a vitamin K food composition dataset that is representative of foods consumed in Australia.

**Keywords:** Australia; food composition; menaquinone; phylloquinone; vitamin K

# 1. Introduction

Vitamin K is the family name for a group of fat-soluble compounds, with phylloquinone (PK; also known as vitamin K1) and menaquinones (MKs) as the major compounds. PK is found in vegetables and plant oils, while MKs are found in meat, eggs, and fermented foods, such as dairy (Schurgers & Vermeer, 2000). Recently, the focus on vitamin K has shifted from a single-function vitamin for normal blood coagulation towards a multi-function vitamin, with an essential role in maintaining normal bone, neurological and vascular function (Halder et al., 2019).

Vitamin K is a nutrient of concern in North America and Europe, where intakes are low - often below the requirements for normal blood coagulation and insufficient for optimal health (Hayes et al., 2016; McCann & Ames, 2009; Thane, Paul, Bates, Bolton-Smith, Prentice, & Shearer, 2002; Turck et al., 2017; Wallace, McBurney, & Fulgoni, 2014). Since there is no single biomarker for vitamin K status, assessing population vitamin K status is challenging and costly; hence, estimating dietary intakes to infer status is a convenient alternative (Shea & Booth, 2016). In the absence of sufficient dose-response data to set an Estimated Average Requirement (EAR) for vitamin K, varying Adequate Intake (AI) recommendations have been made in Australia, Europe and the US. An AI is set at 60 and 70 µg /day for Australian women and men, respectively (National Health and Medical Research Council, 2014), and at 70 µg /day for all European adults (Turck et al., 2017). The AI is higher in the US at 90 and 120 µg /day for women and men, respectively (Institute of Medicine US) Panel on Micronutrients, 2001).

There is currently no authoritative source of vitamin K composition data, particularly for MKs (Shea et al., 2016), for accurately estimating vitamin K intakes. Various countries, including the US, UK, Denmark, the Netherlands and New Zealand, have PK composition data in their national food composition databases. There is growing interest in MKs (Shea et al., 2016); however, no country has comprehensive data on the content of MKs in food. Limited data on MKs are available in databases from the US (U.S. Department of Agriculture - Agricultural Research Service, 2019), UK (Food Standards Agency UK, 2008) and The Netherlands (Rijksinstituut voor Volksgezondheid en Milieu (RIVM), 2019), and in published studies for a small number of foods (Elder, Haytowitz, Howe, Peterson, & Booth, 2006; Fu et al., 2017; Fu, Shen, Finnan, Haytowitz, & Booth, 2016; Jensen, Daugintis, & Jakobsen, 2021; Jensen, Ložnjak Švarc, & Jakobsen, 2021; Karl, Fu, Dolnikowski, Saltzman, & Booth, 2014; Koivu-Tikkanen, Ollilainen, & Piironen, 2000; Palmer et al., 2021; Schurgers et al., 2000; Vermeer, Raes, van't Hoofd, Knapen, & Xanthoulea, 2018). Using UK food composition tables and published studies, preliminary estimates of dietary vitamin K intakes in the Irish population showed that intakes of MKs were of similar magnitude to intakes of PK (Kingston et al., 2019). This suggests that measuring both PK and MKs in foods is important for accurate reporting of vitamin K intakes.

Vitamin K remains largely unexplored in the Australian population: there is no estimate of vitamin K intakes as there have been no vitamin K composition data for Australian foods until recently. A preliminary database has been produced for selected foods available in Australia for PK, MK-4 and MK-7 (Palmer et al., 2021); however, nationally representative data from larger sample sizes across a greater range of foods and K vitamers are still required. Hence, the aim of this study was to explore the content of eight K vitamers (PK and MK-4 to -10) in cheese, yoghurt and meat products sourced from multiple cities across Australia.

**2. Materials and methods**

*2.1 Sampling and sample preparation*

Primary samples ($n = 288$) of cheese, yoghurt, and meat products were purchased in Sydney ($n = 90$), Melbourne ($n = 114$) and Perth ($n = 84$) as part of a larger sampling program for vitamin D described elsewhere (Dunlop et al., 2021). These three cities represent both the east and west coasts of Australia and are where approximately half of Australia's population resides and purchases food.

Samples were purchased from supermarkets and specialty shops, including independent butchers across a 10-month period (August 2018-June 2019) in order to capture any potential seasonal variation. The location and date of purchase and weight of samples were recorded. Samples were kept chilled from the time of purchase to preparation. They were packaged such that they were protected from heat and light and that any liquid contents were contained during transportation. Samples were transported to the National Measurement Institute of Australia (NMI), Melbourne, for preparation. Foods were prepared and cooked as they would usually be consumed, omitting oil and other ingredients, except for small amounts of water when needed to prevent adherence to cooking vessels. For each city in which a food was sampled, six samples were purchased. Each group of six primary samples of the same food type purchased in the same city were combined, using equal aliquots, into homogenized composite samples for analysis (total $n$ = 48; Sydney $n$ = 15; Melbourne n = 19; Perth $n$ = 14; Table 1).

Immediately after preparation, the composite samples were stored frozen at -20°C, protected from light and oxygen to prevent loss of the K vitamers during the storage period (Indyk, Shearer, & Woollard, 2016). In July 2021, the frozen samples were packed into thermal boxes with sufficient dry ice to ensure that samples remained frozen from collection to delivery and were couriered by fastest available means (three-day transit time) to the Technical University of Denmark (DTU), Lyngby, Denmark, for analysis of K vitamers.

*2.2 Analysis*

Fat was measured in duplicate at NMI following sample preparation using either Soxhlet (Food Science Australia, 1998) or Mojonnier extraction (AOAC International, 2005). The K vitamers, PK and MK-4 to MK-10, were analysed at DTU in duplicate or triplicate using a validated method described in detail previously (Jäpelt & Jakobsen, 2016; Jensen, Ložnjak Švarc, et al., 2021; Jensen, Rød, Ložnjak Švarc, Oveland, & Jakobsen, 2022). Briefly, all analytical procedures were conducted under yellow light or by use of amber glassware or foil coverings throughout the process. Between 0.3 and 0.5 g of sample, depending on the likely content of vitamin K, was combined with an internal standard (IS) mix. The IS mix provided 125 ng each of labelled IS PK-[$^2$H$_7$] (d7- PK), MK-4-[$^2$H$_7$] (d7-MK-4), MK-7-[$^2$H$_7$] (d7-MK-7), and MK-9-[$^2$H$_7$] (d7-MK-9) (IsoSciences, Ambler, PA). The vitamin K and IS were then extracted using 2-propanol, n-heptane and water. Following extraction from the food matrix, extracts underwent lipase treatment, using the enzymes Lecitase™ Ultra and Lipozyme® TL 100L (Novozymes, Bagsværd, Denmark), followed by an extraction of the vitamin K vitamers and the IS. The lipase treatment and extraction processes were repeated once. A solid phase extraction (SPE) clean-up using a silica column was then carried out, after which the sample was transferred to a vial.

Calibration standards were prepared with 250 ng/mL each of labelled standards, d7-PK, d7-MK-4, d7-MK7 and d7-MK-9 and 2.5, 5, 10, 25, 50, 100, 250, 375 or 500 ng/mL unlabeled standards (PK, MK-4, MK-7 and MK-9; Sigma Aldrich, Darmstadt, Germany) dissolved in ethanol. Calibration standards and samples were analysed using UHPLC (1290 Infinity II, Agilent Technologies, Santa Clara, CA) coupled with Ascentis® Express C18 columns, (5 mm + 10 cm) x 2.1 mm, 2.7 µm; Supelco, Bellefonte, PA) and connected to the Triple Quadrupole MS (6470, Agilent Technologies, Santa Clara, CA). For quantification of PK, MK-4, MK-7 and MK-9 the respective four IS was used. The calibration curves of MK-4, MK-7, MK-9 and MK-9 in combination with calibration factors were used to quantify the content of MK-5, MK-6, MK-8 and MK-10, respectively, as described previously (Jensen et al., 2022).

*2.3. Quality assurance of the analytical method*

Trueness for PK was confirmed by analysing certified reference materials (Kelp 3232 and Infant formula 1849, NIST, Gaithersburg, ML). To assess trueness and precision during the analytical run, house reference materials (hard cheese and blue cheese) containing PK, MK-4 and MK-6 to MK-9, were analysed in each analytical run.

*2.3 Data handling of the results*

For the composite samples, duplicated results were averaged to provide concentrations for each K vitamer in each food for each city in which it was purchased. Concentrations of K vitamers were averaged across products purchased in each city, and then across cities to produce national average concentrations. For foods with one analytical sample, we reported concentrations as the mean of duplicated or triplicated analyses. For foods with two or more analytical samples, concentrations were reported as the mean of duplicated or triplicated analyses across cities ± standard deviation.

**3. Results**

*3.1. Analytical quality assurance results*

Limits of quantification were 0.5, 0.5, 1, 1, 2.5, 5, 1 and 5 µg/100 g for PK, MK-4, MK-5, MK-6, MK-7, MK-8, MK-9 and MK-10, respectively. The results for the certified reference materials (Kelp (NIST3232) 442 ± 43 PK ng/g (n=4); Infant Formula (NIST1849) 2280 ± 210 ng PK/g (n=4)) were within the certified ranges of 434 ± 81 ng PK/g and 2200 ± 180 ng PK/g, respectively. The trueness for the values of the MKs are reported elsewhere, by comparing to a different analytical method (Jensen et al., 2022). The precision achieved in our study was 11% for PK (n=28), 14% for MK-4 (n=46), and 22% for MK-9 (n=8). For the remaining MKs, the precision was previously assessed as <25% (Jensen et al., 2022).

*3.2 Analytical results*

We found PK in all samples except bacon, chicken, ham and pork (Table 2). The greatest concentration of PK was found in lamb liver. MK-4 was quantified above the LOQ (0.5 µg /100 g) in 46 of the samples, with the greatest concentrations found in chicken, particularly leg meat (with skin). MK-5 was found in Cheddar cheese and cream cheese only, with similar concentrations. MK-8 was found in Cheddar cheese only. MK-9 was found in all cheese products, including cheesecake; the concentration in Cheddar cheese was considerably greater than in other cheese products. Neither MK-6, MK-7 nor MK-10 were quantified in any samples.

**4. Discussion**

This study provides an insight into the vitamin K content of Australian cheese, yoghurt and meat products, sampled across multiple cities and seasons. At least one K vitamer was found in each sample, and some foods were useful sources of vitamin K, particularly PK, MK-4 and MK-9.

When establishing new values for the vitamin K content of foods, the validation of the analytical method is essential. The challenge in the method validation of the MKs is the lack of certified reference materials. For the first time, to justify the trueness of our method, we included a comparison to a method using post-column derivatisation followed by fluorescence detection for PK and MKs (Jensen et al., 2022). Few others have reported results for PK and MKs in foods (Elder et al., 2006; Fu et al., 2017; Fu et al., 2016; Jensen, Daugintis, et al., 2021; Jensen, Ložnjak Švarc, et al., 2021; Karl et al., 2014; Koivu-Tikkanen et al., 2000; Palmer et al., 2021; Schurgers et al., 2000; Vermeer et al., 2018). Reported LOQs (per 100 g) for LC-fluorescence, LC-APCI-MS and LC-ESI-MS/MS methods range from 0.05-1.4 µg for PK, 0.1-4.0 µg for MK-4, 0.1-0.7 µg for MK-5, 0.1-0.6 µg for MK-6, 0.1-2.6 µg for MK-7, 0.1-4.3 µg for MK-8, 0.1-2.4 µg for MK-9 and 0.1-4.0 µg for MK-10 (Elder et al., 2006; Jensen et al., 2022; Karl et al., 2014; Koivu-Tikkanen et al., 2000; Palmer et al., 2021). Thus, our LOQs are comparable with those that have been obtained by others.

In our study, where the majority of cheeses included were made from ≥ 95% ingredients of Australian or New Zealand origin, we found PK, MK-4 and MK-9 in all cheese products, while Cheddar cheese also contained MK-5 and MK-8. The predominant vitamer in all included cheese products was MK-4. We found greater concentrations of PK and MK-4 in Cheddar and Camembert/Brie compared to an earlier Australian study; however, in that study only three samples of each variety were included and only PK, MK-4 and MK-7 were measured (Palmer et al., 2021). We found that MK-9 contributed a considerable proportion of vitamin K content in Cheddar and Brie/Camembert varieties, while Cheddar also contained reasonable concentrations of MK-5 and MK-8. Previous studies have been conducted at DTU to measure the vitamin K content of selected cheeses purchased in Denmark (Jensen, Daugintis, et al., 2021; Jensen, Ložnjak Švarc, et al., 2021). MK-4 was the predominant vitamer found in Mozzarella in both our study and a Danish study (Jensen, Ložnjak Švarc, et al., 2021); however, the K vitamer profiles differed in that MK-9 (1.2 ± 0.6 µg/100 g) was found in our Mozzarella samples, but not in the earlier study of Mozzarella purchased in Denmark. In earlier Danish studies, MK-9 was the predominant K vitamer in Gouda, Tistrup, Cheddar and Danablue varieties, all of which also contained MK-7 (1.0-3.2 µg/100 g) (Jensen, Daugintis, et al., 2021; Jensen, Ložnjak Švarc, et al., 2021). MK-7 was not quantified in any of the cheese products included in our study.

Variation in K vitamer concentration and profile in cheeses has also been seen in studies conducted elsewhere. A Dutch study found PK (0.3-10.4 µg/100 g) and MK-4 to MK-9 (0.1-51.1 µg/100 g) in hard, soft and curd cheeses (Schurgers et al., 2000). Similarly, a more recent study measured PK and MK-4 to MK-10 in cheeses (Gouda, Milner, Slankie, Edam, Maasdam and curd cheeses) purchased in The Netherlands, finding quantifiable concentrations of PK and MK-4 to MK-9 in all samples except Edam, in which only PK, MK-4, MK-8 and MK-9 were quantified (Vermeer et al., 2018). A study conducted in Finland found that Edam cheese contained quantifiable concentrations of PK (1.9 ± 1.3 µg/100 g) and MKs 4-10 (0.5-30.0 µg/100 g), while Emmental contained quantifiable concentrations of PK (2.6-3.0 µg/100 g) and MK-4 (5.2-6.1 µg/100 g), with traces of MK-6 and MK-7 (Koivu-Tikkanen et al., 2000). More recently, in the US, K vitamers have been measured in a wider range of processed, fresh (goat, feta, ricotta, Cotija, cottage and Mozzarella), blue (Gorgonzola and other blue cheeses), soft (Brie, Camembert, crème fraiche, Limburger, mascarpone), semi-soft (Monterey Jack, Havarti, Fontina, Gouda, Swiss and cream cheese) and hard (Cheddar – regular and full fat - and Parmesan) cheeses (Fu et al., 2017). In that study, all regular-fat cheese contained, PK, MK-4, and MK-7 to MK-11. Some regular-fat varieties also contained MK-5, MK-6, MK-12 and MK-13; however, reduced-fat cottage and Cheddar cheeses did not; there was also no PK or MK-4 detected in reduced fat cottage cheese (Fu et al., 2017). In contrast to our study, MK-9, rather than MK-4, was the predominant vitamer in all cheese varieties sampled in that study (Fu et al., 2017).

The presence of the longer-chain MKs in fermented dairy products, such as cheese, has been suggested to be due to use of bacterial cultures in the fermentation process (Schurgers et al., 2000; Shearer, 1997), with the variability seen in MK profiles being due to use of different microbial species in different production methods (Fu et al., 2017). However, the starter culture used, water content, fat content and ripeness of cheese were investigated in a recent Danish study, with the finding that none of these factors explained K vitamer content in five varieties (Brie, Cheddar, Danablu, Hirtenkäse and Danbo) studied (Jensen, Daugintis, et al., 2021).

We found small amounts of PK (0.5 ± 0.0 µg/100 g) and reasonable amounts of MK-4 (1.8 ± 0.3 µg/100 g) in full-fat yoghurt, but only small amounts of MK-4 (0.2 ± 0.3 µg/100 g) in low-fat yoghurt. No other K vitamers were quantitated in our yoghurt samples. Our findings appear similar to those of an earlier Australian study (Palmer et al., 2021); however, making a comparison is difficult as fat content was not reported in the other study. A US study had contrasting results, with various K vitamers (PK and MK-4,MK-9 to MK-11) found in full-fat regular (mean fat 4.6%) and Greek (mean fat 4%) yoghurts and none detected in fat-free versions of those yoghurt varieties (Fu et al., 2017). Across a range of dairy milk, yogurts, kefirs, cream and cheeses, overall vitamin K content was found to be proportional to fat content (Fu et al., 2017). In that study, the greatest concentrations were of MK-9 in both full-fat varieties (mean = 13.2 µg/100 g in regular yoghurt and 14.8 µg/100 g in Greek yoghurt) (Fu et al., 2017). Studies conducted in Finland (Koivu-Tikkanen et al., 2000) and France (Manoury, Jourdon, Boyaval, & Fourcassié, 2013) have also found a wide range of MKs, dominated by MK-9 in fermented and soured milk products; however, plain yoghurt (2.5% fat) analysed in the Finnish study contained only PK, MK-4 and MK-5 (Koivu-Tikkanen et al., 2000). In the Netherlands, concentrations of MK-4, MK-5 and MK-8 were quantified in whole milk yoghurt, and MK-8 in skimmed milk yoghurt; MK-9 was not detected in either variety (Schurgers et al., 2000).

We found a distinct difference in K vitamer profiles between ruminant animal products (beef and lamb products) that contained PK and MK-4 and mono-gastric animal products (chicken and pork products) that contained MK-4 only. Generally, ruminant animal products contained greater concentrations of MK-4 than PK; however, the concentration of PK in lamb liver was greater than that of MK-4. Our results for PK in most meat products were similar to those from another Australian study (Palmer et al., 2021). Compared to that study, mean concentrations of MK-4 in our study were 2.6-24.8 µg/100 g lower in beef, pork, ham and salami and 7.5 µg/100 g higher in beef sausage. These differences may be due to the timing, location and breadth of sampling. For our study, we developed a national sampling plan to capture differences across region and season. We also sampled different cuts of meat, and we prepared and cooked foods as they would be consumed in the home, eliminating the need for conversion factors for raw foods.

There is considerable variation in the vitamin K content of meat products in other countries. Samples of Dutch beef contained both PK (0.6 µg/100 g) and MK-4 (1.1 µg/100 g) (Schurgers et al., 2000), while in the US, beef steak contained 1.9 µg/100 g MK-4, but no PK (U.S. Department of Agriculture - Agricultural Research Service, 2019). In the UK, a much greater concentration of 7.2 µg/100 g PK was found in beef mince, with a lesser concentration found in roast beef (0.2 µg/100 g) (Food Standards Agency UK, 2008). Reasonable mean concentrations of 8.5-8.9 µg/100 g MK-4 were found in chicken from the Netherlands (Schurgers et al., 2000); however, this is considerably less that the MK-4 concentrations found in our Australian chicken samples. Elsewhere, a number of K vitamers have been quantified in pork products. In the Netherlands, four K vitamers (PK and MK-4, MK-7 and MK-8) were found in pork steak, and salami (which commonly contains both pork and beef) contained PK and MK-4 (Schurgers et al., 2000). A US study found PK and MK-4, MK-10 and MK-11 in pork sausage and cooked Canadian bacon (Fu et al., 2016). In pork products, we only found MK-4, which may be due to differing production methods (e.g., fermentation methods).

Collectively, these studies indicate that K vitamer concentration and profile can vary considerably within and between food varieties and by geographic location. The MK-4 found in animal produce may be a product of conversion of PK or menadione, obtained from the animal's diet (Fu et al., 2016; Hirota et al., 2013), to MK-4 (Booth, 2012; Schurgers et al., 2000; Thijssen, Drittij-Reijnders, & Fischer, 1996). As menadione is the predominant form of vitamin K in feed products used in many systems of animal husbandry (Booth, 2012; Fu et al., 2016; Thijssen et al., 1996), it is considered a likely source of MK-4 in farmed animals that receive menadione-rich diets (Fu et al., 2016). Therefore, the vitamin K content of produce may vary by location and based on the production methods used (e.g., livestock feed profile and the availability of other natural sources of vitamin K in the local environment, such as grasses).

Further work is needed to allow estimation of vitamin K intakes in Australia. Although population intakes of most nutrients were estimated from the 2011-2013 Australian Health Survey (Australian Bureau of Statistics, 2014) and the 2011-2013 Australian Aboriginal and Torres Strait Islander Health Survey (Australian Bureau of Statistics, 2015), vitamin K intakes could not be quantified due to the lack of vitamin K composition data. Borrowing data from international food composition databases is not appropriate: environment and production methods in Australia may impact on the vitamin K content of foods, and values in international food composition databases are often borrowed from other databases or are derived from unknown sources. Since vitamin K intake may vary by region, perhaps due to dietary patterns and the types of foods favoured by different populations (Booth, 2012), locally-relevant food composition data and intake estimates are needed. Furthermore, some foods (e.g., kangaroo, emu) are specific to the Australian population. Hence, to estimate vitamin K intakes in the Australian population, Australia-specific vitamin K composition data are required. National food sampling is required for data to be of sufficient quality for use in national food composition tables and for use in estimating usual intakes.

We used innovative methods developed and validated by our team to measure PK and MKs in foods (Jäpelt et al., 2016; Jensen, Ložnjak Švarc, et al., 2021). These methods are rapid, highly sensitive and specific, with the capacity to detect low levels of PK and MKs in a range of complex food matrices in a single analytical run, hence accommodating high and efficient throughput. The capacity to measure even small amounts of vitamin K in foods is important, since some food sources of vitamin K are widely consumed, and small levels of nutrients are cumulatively significant across the diet. We measured PK and MK-4 to MK-10 concentrations in all foods sampled. While samples were stored frozen for 2-3 years, this is expected to have had minimal impact on K vitamer content (Indyk et al., 2016). Our sampling plan was carefully designed to account for potential geographical and seasonal variations across the Australian continent. However, a general limitation of food composition data is that they may not represent the precise nutrient content of individual consumed foods.

This study provides new data for vitamin K in cheese, yoghurt and meat products sourced in Australia. All samples contained at least one K vitamer. Our study contributes to the limited vitamin

K composition data available in Australia and globally, and adds to the growing evidence that the K vitamer profile and concentration of foods can vary greatly by region. A larger project is needed to develop a comprehensive analytical food composition database for K vitamers (PK and MKs) in a wide range of foods commonly consumed in Australia so that intakes of vitamin K can be estimated in the population.

**Abbreviations:**

| | |
|---|---|
| AI | Adequate Intake |
| EAR | Estimated Average Requirement |
| DTU | Danish Technical University |
| HPLC | high performance liquid chromatography |
| LOQ | limit of quantitation |
| MK | menaquinone |
| NIST | National Institute of Standards and Technology |
| NMI | National Measurement Institute of Australia |
| PK | phylloquinone |
| RPD | relative percent difference |

**Declarations of interest:** none

**Author contributions: Eleanor Dunlop:** Data curation, Writing – Original draft, Project administration. **Jette Jakobsen**: Methodology, Investigation, Writing – Review and editing. **Marie Bagge Jensen:** Methodology, Investigation, Writing – Review and editing. **Jayashree Arcot:** Writing – Review and editing. **Liang Qiao:** Writing – Review and editing. **Judy Cunningham:** Methodology, Writing – Review and editing. **Lucinda J Black:** Conceptualization, Funding acquisition, Methodology, Project administration, Writing – Review and editing.

**Funding:**

Sample collection was funded by the National Health and Medical Research Council as part of GNT1140611. Sample analysis was funded by the Western Australian Department of Health. LJB is supported by MS Western Australia (MSWA), an MS Australia Postdoctoral Fellowship and a Curtin University Research Fellowship.

**Table 1.** Purchase location and preparation of cheese, yoghurt and meat products purchased in Australia

| Sample description | Primary samples, *n* | Purchase location(s) | Preparation | Country of origin |
|---|---|---|---|---|
| Cheese, cheddar | 18 | Sydney, Melbourne, Perth | None | All samples: 96-99+% ingredients of Australian or New Zealand origin |
| Cheese, feta | 6 | Sydney | None | 3 samples: Australia<br>2 samples: Denmark<br>1 sample: Greece |
| Cheese, mozzarella | 12 | Sydney, Perth | None | All samples: 95-99% ingredients of Australian or New Zealand origin |
| Cheese, brie or camembert | 6 | Melbourne | None | 5 samples: Australia<br>1 sample: Denmark |
| Cheese, cream cheese, regular fat | 6 | Melbourne | None | All samples: 95-100% ingredients of Australian origin |
| Cheesecake, plain or flavoured | 6 | Melbourne | None | 2 samples: 47-74% ingredients of Australian origin<br>4 samples: origin unknown |
| Yoghurt, flavoured or added fruit, full fat (3-5% fat) | 18 | Sydney, Melbourne, Perth | None | All samples: 83-99% ingredients of Australian origin |
| Yoghurt, flavoured or added fruit, reduced fat (1-2% fat) | 18 | Sydney, Melbourne, Perth | None | All samples: 90-99% ingredients of Australian origin |
| Beef mince, regular fat | 18 | Sydney, Melbourne, Perth | Pan fried without oil | Australia |
| Beef, steak, semi-trimmed | 18 | Sydney, Melbourne, Perth | External fat removed, grilled | Australia |
| Beef sausage | 18 | Sydney, Melbourne, Perth | Grilled/BBQ/pan fried | 13 samples: 92-99% ingredients of Australian origin<br>5 samples: origin unknown |
| Chicken, leg meat with skin | 18 | Sydney, Melbourne, Perth | Baked | Australia |
| Chicken, skinless breast fillets | 18 | Sydney, Melbourne, Perth | Pan fried without oil | Australia |
| Kangaroo, steak | 6 | Melbourne | Pan fried without oil | Australia |
| Lamb, chops, semi-trimmed, | 18 | Sydney, Melbourne, Perth | Grilled | Australia |
| Liver, lamb | 6 | Melbourne | Pan fried without oil | Australia |
| Lard and dripping | 6 | Melbourne | None | 5 samples: Australia<br>1 sample: 95% ingredients of Australian origin |
| Pork Chops, semi-trimmed | 18 | Sydney, Melbourne, Perth | Grilled/BBQ | Australia |
| Pork, minced | 18 | Sydney, Melbourne, Perth | Pan fried without oil | Australia |
| Bacon, partly trimmed | 18 | Sydney, Melbourne, Perth | Pan fried without oil | 12 samples: 10-21% ingredients of Australian origin<br>6 samples: origin unknown |
| Ham, sliced | 12 | Sydney, Perth | None | Australia, North America, Europe |
| Salami, regular fat | 6 | Melbourne | None | All samples: 92-100% ingredients of Australian origin |

3 **Table 2.** PK and MK -4 to -10 content of cheese, yoghurt and meat products purchased in Australia

| Product | Primary samples (*n*) | Analytical samples (*n*) | Fat g/100 g | PK | MK-4 | MK-5 | MK-6 | MK-7 | MK-8 | MK-9 | MK-10 |
|---|---|---|---|---|---|---|---|---|---|---|---|
| | | | | | | | (µg/100 g) | | | | |
| Cheese, Cheddar | 18 | 3 | 32.9 | 2.79 ± 0.32 | 9.12 ± 2.08 | 1.05 ± 0.30 | < LOQ | < LOQ | 3.95 ± 2.19 | 8.07 ± 1.84 | < LOQ |
| Cheese, feta | 6 | 1 | 19.7 | 1.92 | 6.33 | < LOQ | < LOQ | < LOQ | < LOQ | 4.16 | < LOQ |
| Cheese, Mozzarella | 12 | 2 | 22.1 | 1.86 ± 0.42 | 5.54 ± 1.00 | < LOQ | < LOQ | < LOQ | < LOQ | 1.25 ± 0.55 | < LOQ |
| Cheese, Brie or Camembert | 6 | 1 | 32 | 2.95 | 24.23 | < LOQ | < LOQ | < LOQ | < LOQ | 5.56 | < LOQ |
| Cheese, cream cheese, regular fat | 6 | 1 | 30 | 3.02 | 11.52 | 0.95 | < LOQ | < LOQ | < LOQ | 1.81 | < LOQ |
| Cheesecake, plain or flavoured | 6 | 1 | 15.1 | 2.96 | 6.20 | < LOQ | < LOQ | < LOQ | < LOQ | 1.67 | < LOQ |
| Yoghurt, flavoured or added fruit, full fat | 18 | 3 | 4.4 | 0.54 ± 0.04 | 1.75 ± 0.32 | < LOQ | < LOQ | < LOQ | < LOQ | < LOQ | < LOQ |
| Yoghurt, flavoured or added fruit, reduced fat | 18 | 3 | 1.4 | < LOQ | 0.17 ± 0.26 | < LOQ | < LOQ | < LOQ | < LOQ | < LOQ | < LOQ |
| Beef mince, regular fat | 18 | 3 | 18.4 | 2.01 ± 0.32 | 8.16 ± 2.14 | < LOQ | < LOQ | < LOQ | < LOQ | < LOQ | < LOQ |
| Beef steak, semi-trimmed | 18 | 3 | 5.9 | 0.80 ± 0.12 | 2.44 ± 0.50 | < LOQ | < LOQ | < LOQ | < LOQ | < LOQ | < LOQ |
| Beef, sausage | 18 | 3 | 19.2 | 3.80 ± 0.86 | 10.47 ± 3.01 | < LOQ | < LOQ | < LOQ | < LOQ | < LOQ | < LOQ |
| Chicken, leg meat with skin | 18 | 3 | 8.4 | < LOQ | 58.16 ± 8.59 | < LOQ | < LOQ | < LOQ | < LOQ | < LOQ | < LOQ |
| Chicken, skinless breast fillets | 18 | 3 | 2.7 | < LOQ | 26.78 ± 4.81 | < LOQ | < LOQ | < LOQ | < LOQ | < LOQ | < LOQ |
| Kangaroo steak | 6 | 1 | 2.4 | 0.84 | 1.43 | < LOQ | < LOQ | < LOQ | < LOQ | < LOQ | < LOQ |
| Lamb, chops, semi-trimmed | 18 | 3 | 13.3 | 1.40 ± 0.89 | 12.44 ± 2.64 | < LOQ | < LOQ | < LOQ | < LOQ | < LOQ | < LOQ |
| Liver, lamb | 6 | 1 | 9.3 | 4.93 | 2.56 | < LOQ | < LOQ | < LOQ | < LOQ | < LOQ | < LOQ |
| Lard dripping | 6 | 1 | 99.4 | 4.00 | 10.44 | < LOQ | < LOQ | < LOQ | < LOQ | < LOQ | < LOQ |
| Pork chops, semi-trimmed | 18 | 3 | 6.5 | < LOQ | 4.99 ± 2.28 | < LOQ | < LOQ | < LOQ | < LOQ | < LOQ | < LOQ |
| Pork, minced | 18 | 3 | 14.6 | < LOQ | 8.08 ± 1.69 | < LOQ | < LOQ | < LOQ | < LOQ | < LOQ | < LOQ |
| Bacon, partly trimmed | 18 | 3 | 18.3 | < LOQ | 16.10 ± 3.08 | < LOQ | < LOQ | < LOQ | < LOQ | < LOQ | < LOQ |
| Ham, sliced | 12 | 2 | 3.7 | < LOQ | 3.92 ± 0.92 | < LOQ | < LOQ | < LOQ | < LOQ | < LOQ | < LOQ |
| Salami, regular fat | 6 | 1 | 29.2 | 1.75 | 15.61 | < LOQ | < LOQ | < LOQ | < LOQ | < LOQ | < LOQ |

4 Concentrations are presented as the mean of duplicated or triplicated analyses for foods with one analytical sample. For foods with two or more analytical samples, concentrations are
5 presented as the mean of duplicated or triplicated analyses across cities ± standard deviation
6 LOQ = 0.5, 0.5, 1, 1, 2.5, 5, 1 and 5 µg/100 g for PK, MK-4, MK-5, MK-6, MK-7, MK-8, MK-9 and MK-10, respectively
7 LOQ, limit of quantitation; MK, menaquinone; PK, phylloquinone